\documentclass[a4paper]{report}
\usepackage[utf8]{inputenc}
\usepackage[T1]{fontenc}
\usepackage{RJournal}
\usepackage{amsmath,amssymb,array}
\usepackage{booktabs}

\begin{document}

%% do not edit, for illustration only
\sectionhead{Contributed research article}
\volume{XX}
\volnumber{YY}
\year{20ZZ}
\month{AAAA}

%% replace RJtemplate with your article
\begin{article}

% !TeX root = RJwrapper.tex
\title{An Overview on the Landscape of {R} Packages for Credit Scoring}
\author{by Gero Szepannek}

\maketitle

\abstract{
The credit scoring industry has a long tradition of using statistical tools for loan default probability prediction and domain specific standards have been established long before the hype of machine learning. Although several commercial software companies offer specific solutions for credit scorecard modelling in {R} explicit packages for this purpose have been missing long time. In the recent years this has changed and several packages have been developed with an dedicated to credit scoring.

The aim of this paper is to give a structured overview on these packages. This may guide users to select the appropriate functions for a desired purpose further hopefully will help reducing redundant development activities in the future. The paper is guided by the chain of subsequent modelling steps as they are forming the typical scorecard development process. 
}

\section{Introduction}\label{sec:introduction}
In the credit scoring industry there is a long tradition of using statistical tools for loan default probability prediction and domain specific standards have been established long before the hype of machine learning. Although several commercial software companies such as e.g. SAS offer specific solutions for credit scorecard modelling\footnote{\url{https://www.sas.com/en_us/software/credit-scoring.html}} in {R} explicit packages for this purpose have been missing long time and in the CRAN task view on %\emph{Empirical} 
\ctv{Finance} the explicit topic of scorecard modelling is not covered. A ''Guide to Credit Scoring in {R}'' can be found among the CRAN contributed documentations \citep{sharma:2009} being rather dedicated to describing the application of different (binary) classification algorithms to credit scoring data than to emphasize the common subsequent modelling stages as they are typical for scorecard modelling processes. This can be a result of the circumstance that at that time no explicit packages where available in {R} for solving this kind of task.

In the recent years this has changed and several packages have been submitted to CRAN with an explicit scope of credit risk scorecard modelling, such as \CRANpkg{scorecard} \citep{scorecard}, \CRANpkg{scorecardModelUtils} \citep{scorecardModelUtils}, \CRANpkg{smbinning} \citep{smbinning}, \CRANpkg{woeBinning} \citep{woeBinning}, \CRANpkg{woe} \citep{woe}, \CRANpkg{woeR} \citep{woeR},  \CRANpkg{Information} \citep{Information}, \CRANpkg{InformationValue} \citep{InformationValue}, \CRANpkg{glmdisc} \citep{glmdisc}, \CRANpkg{glmtree} \citep{glmtree}, \CRANpkg{Rprofet} \citep{Rprofet}, and \CRANpkg{boottol} \citep{boottol}. 

\begin{figure}[htbp]
  \centering
  \includegraphics[width = 14cm]{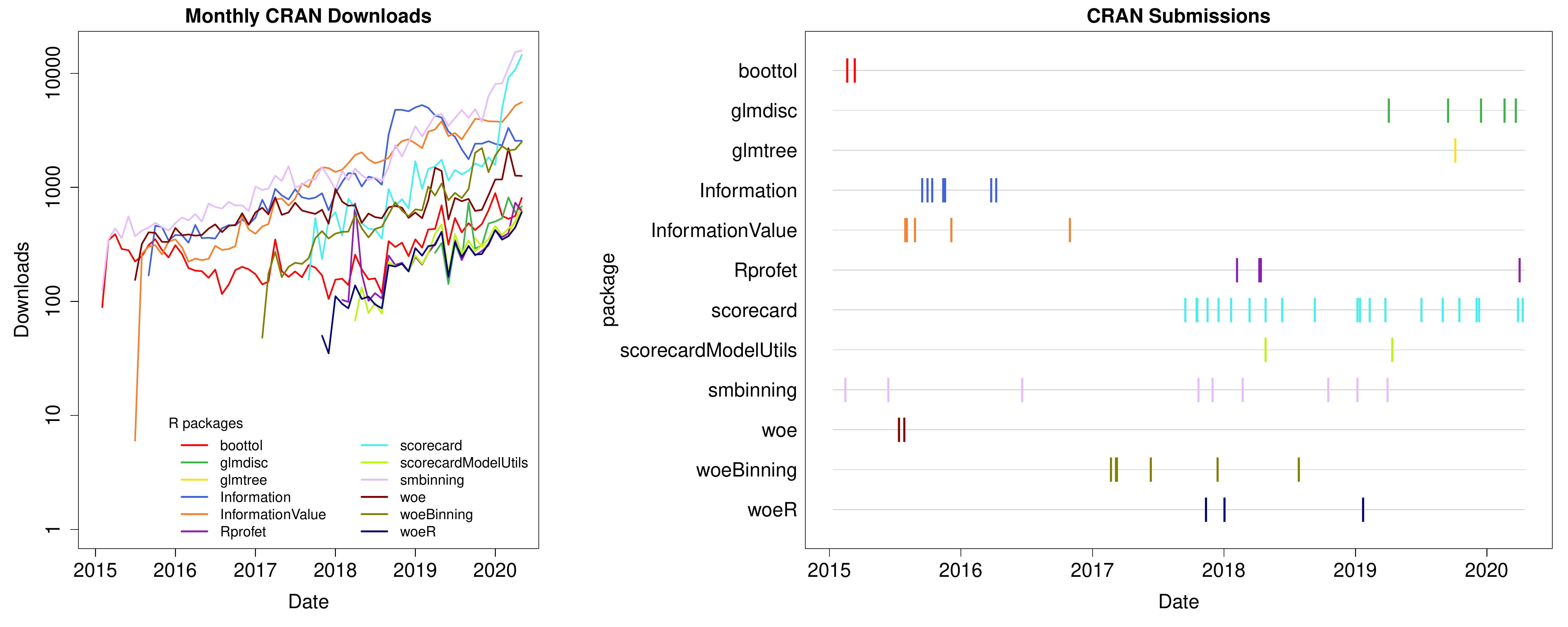}
  \caption{CRAN release activity and download statistics \citep[as returned by \CRANpkg{cranlogs},][]{cranlogs} of packages avialable on CRAN.}
    \label{figure:pkgstats}
\end{figure}

Figure~\ref{figure:pkgstats} gives an overview of the packages and their popularity in terms of the number of their CRAN downloads as well as their activity and existence as observable by their CRAN submission dates. It can be seen that the packages \pkg{smbinning} \pkg{InformationValue} and  \pkg{Information} are the oldest and most popular ones. The packages \pkg{scorecard} and \pkg{smbinning} are most regularly updated.  

In addition, some packages are available on Github but not on CRAN such as \pkg{scoringTools} \citep{scoringTools}, \pkg{riskr} \citep{riskr} and \pkg{creditR} \citep{creditR}. As all of these packages have arisen during the last five years the aim of this paper is to give a structured overview on existing packages. On one hand it may guide users to selection of the appropriate functions for a desired purpose by listing pros and cons of existing functions and on the other hand it hopefully will help reducing redundant development activities by programming multiple solutions for one and the same task and finally maybe even improving existing methodolgy by working out advantages as they are covered by different packages. 

The typical steps in credit risk scorecard modelling are very closely related to the general process definition for data mining as it is given by KDD, CRISP-DM or SAS's SEMMA \citep[cf. e.g.][]{azevedo:2008}. A detailed description of the subsequent steps can be found in \citet{siddiqi:2006}, \citet{finlay:2012},  \citet{anderson:2019} or \citet{thomas:2002}. It turns out that strong emphasis is laid on possibilities for manual intervention after each modelling step. Therefore functions to summarize and visualize the intermediate results of each single step are of great importance.

The presentation of the package landscape will be guided by these steps where one Section will be dedicated to each stage. In each Section the available packages will be presented together with their advantages and disadvantages. The typical development steps are denoted by:
\begin{enumerate} 
	\item \nameref{sec:binning}
	\item \nameref{sec:preselection}
	\item \nameref{sec:scorecard}
	\item \nameref{sec:performance}
	\item \nameref{sec:ri}
\end{enumerate} 
In addition to the aforementioned packages, some other packages not explicitly designed for scorecard modelling purposes but also provide functionalities that are very helpful in the scorecard modelling context and thus can serve to improve the analyst's life. If available, such functionalities will also be mentioned in the corresponding  Sections.

Note that traditionally, logistic regression is used for credit risk scorecard modelling despite the current hype around modern machine learning methods as they are provided by frameworks such as e.g. \CRANpkg{mlr3} (\citealp{mlr3, lang19}) or \CRANpkg{caret} (\citealp{caret, kuhn:2008}). Studies have investigated potential benefits from using modern machine learning algorithms (\citealp{bischl:2016, szepannek:2017wias, lessmann:2015, baesens:2002, louzada:2016}) but regulators and the General Data Protection Regulation (GDPR) require models to be understandable (cf. e.g. \citealp{fsb:2017, goodman:2017}). The latter issue can be adressed by methodology of \dfn{explainable machine learning}, e.g. using frameworks as provided by the packages \CRANpkg{DALEX} \citep{biecek:2018} or \CRANpkg{iml} \citep{molnar:2017} while taking into account to what extent a model actually is explainable \citep{szepannek:2019arxiv}. It further turned out that the use of current state-of-the-art ML algorithms is not necessarily always beneficial in the credit scoring context (\citealp{chen:2018, szepannek:2017wias}) and should be rather carefully analyzed in each specific situation and interpretable models preferred otherwise \citep{rudin:2019}. For this reason the scope of the paper is dedicated to the traditional way of scorecard modelling as briefly desribed above.

\section{Data}\label{sec:data}

Probably the most common credit scoring data are the {\emph{German Credit Data}} provided by \citet{hoffmann:1994} which are contained in the UCI Machine Learning Repository \citep{dua:2019}. It contains 21 variables: a binary target (\code{creditability}) and 13 categorical as well as seven numeric predictors. The data have 1000 observations in total with 300 defaults (\code{level == "bad"}) and 700 non-defaults (\code{level == "good"}). The data is provided by several {R} packages such as \CRANpkg{klaR} \citep{klaR}, \pkg{woeBinning}, \pkg{caret} or \pkg{scorecard}. For the examples in this paper the data from the \pkg{scorecard} package are used where in addition the levels of the categorical variables such as \code{present.employment.since}, \code{other.debtors.or.guarantors}, \code{job} or \code{housing} are sorted according to their expected order w.r.t. credit risk. 

Other (partly simmulated) example data sets (amongst others loan data of the peer-to-peer lending company Lending Club\footnote{\url{https://www.lendingclub.com/}}) are contained within the packages \pkg{scoringTools}, \pkg{smbinning} and \pkg{riskr} . 

It is common practise to use separate validation data which is not used for model training but only for validation purposes. The manual interventions between the different modelling steps 
do not allow for repetitive resampling strategies such as k-fold cross validation or bootstrapping for model validation as they are e.g. provided by the package \pkg{mlr3} (cf.e.g. Sec.~\nameref{sec:performance}). Instead, usually one single holdout set is used. The package \pkg{scorecard} has a function \code{split\_df()} that splits data according to a pre-specified percentage into training and validation set. For the examples in the remainder of the paper the following data is used:  

\begin{example}
### example 1: load data
library(scorecard)
data(germancredit)
# transform character variable purpose into factor
germancredit$purpose <- as.factor(germancredit$purpose)

tv <- split_df(germancredit, y = "creditability", ratio = c(0.7, 0.3), seed = 42, 
               no_dfs = 2, name_dfs = c("train", "valid"))
train <- tv$train
valid <- tv$valid

# several packages require the target variables to take values 0/1 
train2 <- train; valid2 <- valid
train2$creditability <- as.integer(train2$creditability == "good")
valid2$creditability <- as.integer(valid2$creditability == "good")
\end{example}

Note that some of the packages (\pkg{smbinning}, \pkg{woeR}, \pkg{woe}, \pkg{creditR}, \pkg{riskr}, \pkg{glmdisc}, \pkg{scoringTools} and \pkg{scorecardModelUtils}) do require the target variable to take only values 0 and 1 as in the example's data sets \code{train2} and \code{valid2}. 
Although this is of course easily obtained the package \pkg{scoringModelUtils} contains a function \code{fn\_target()} that does this job and replaces the original target variable by a new one of name \code{Target}.

\section{Binning and weights of evidence}\label{sec:binning}

Binning of numeric variables is often considered to be the most relevant part of a scorecard development. An initial automatic algorithm-based binning is manually checked and -- if necessary -- modified by the analyst variable by variable. On the one hand this is a very time-consuming task but on the other hand this ensures the dependencies between the explanatory variables and the target in the final model to be plausible and helps detecting sampling bias \citep{verstraeten:2005}. Furthermore it allows to model nonlinear dependencies by linear logistic regression in the subsequent \nameref{sec:scorecard} step. The loss of information by aggregation turned out to be comparatively small while this kind of procedure does not take into account for interactions between seveal variables and the target variable \citep{szepannek:2017wias}. The identification of relevant interactions typically needs a lot of business experience and \citet{sharma:2009} suggests to use random forests in order to identify potential interaction candidates.

\noindent {\bf Requirements:} 

It is important to note that binning not just corresponds to exploratory data analysis but its results have to be considered as an integral part of the model, i.e. the resulting preprocessing has to be applied to new data in order to be able to make use of the final scorecard model. For this reason important requirements on an implementation of the binning step are the possibility to \emph{(i) store the binning results} for all variables as well as to \emph{(ii) apply the binning to new data} with some kind of \code{predict()} function.  

The importance of an option to \emph{(iii) manually modify an initial automatic binning} has already been emphasized. This leads to the requirement for a separate function to manipulate a stored binning results object. 
In order to support this \emph{(iv) summary tables} and \emph{(v) visualizations} of the intermediate binning results are helpful. 
In addition, application of binning in practise has to \emph{(vi) deal with missing data or new levels} of categorical variables that did not occur in the training data as e.g. by regulation may be required that holding back information (and the resulting missing values) must not lead to an improvement of the final score. Both should be considered by the implemented binning function.

Often binning is followed by subsequent assignment of numeric \dfn{weights of evidence} to the factor levels $x$ of the binned variable which are given by:
\begin{equation}
	WoE(x) = \log\left(\frac{f(x|y=1)}{f(x|y=0)}\right).
\end{equation}
Note that just like the bins, the WoEs as computed on the training data of course also belong to the model. Further, an implementation of WoE computation has to take into account for potentially occuring bins that are empty w.r.t. the target level $y=0$ (typically by adding a small constant when computing the relative frequencies $f()$). By construction WoEs are linear in the logit of the target variable and thus well suited for subsequent use of logistic regression. The use of WoEs is rather advantageous for small data sets and directly using the bins may increase performance if enough data is available \citep{szepannek:2017wias}. 
On the opposite, using WoEs fixes mononty between the resulting scorecard points and the default rates of the bins, such that only the sign of the monotony has to be checked. 
%The first implementation of weight of evidence computation in {R} had been given by the package \pkg{klaR}. 
It is further usual to associate binned variables with an {\dfn{information value (IV)}}
\begin{equation}
	IV = \sum_x (f(x|1)-f(x|0)) \, WoE(x)
\end{equation}
based on the WoEs which describes the strength of a single variable alone to discriminate between both classes.

\noindent {\bf Available methodology for automatic binning:} 

Several packages provide functions for automatic binning based on {\emph{conditional inference trees}} \citep{hothorn:2006} from the package \CRANpkg{partykit} \citep{partykit}: 
\code{scorecard::woebin()}, \code{smbinning::smbinning()}, \code{scorecardModelUtils::iv\_table()} and \code{riskr::superv\_bin()}. The implementation in the \pkg{scorecardModelUtils} package merges the resulting bins in order to ensure monotonicity in default rates w.r.t. the original variable which might or might not be desired. For the same purpose the package \pkg{smbinning} offers a separate function (\code{smbinning.monotonic()}).
In contrast to all previously mentioned packages, the package \pkg{woeBinning} implements its own tree algorithm where either initial bins of similar WoE are merged (\code{woe.binning()}) 
or the set of bins is binary split (\code{woe.tree.binning()}) as long as the IV of the resulting variables decreases (increases) by a percentage less (more) than a prespecified percentage (argument \code{stop.limit}) while the initial bins are created to be of minimum size (\code{min.perc.total}). 
A similar approach is given by \code{woeR::woe\_binning()} where a prespecified number of initial bins is merged according as long as their corresponding WoEs do not differ by more than a prespecified \code{woe\_cutoff}.

In addition to tree based binning, the \pkg{scorecard} package offers alternative algorithms (argument \code{method}) for automatic binning based on either the $\chi^2$ statistic, or equal width or size of numeric variables.  

An alternative concept for automatic binning is provided by the package \pkg{glmdisc} which is explicitly designed in order to be used in combination with logistic regression modelling for credit scoring \citep{ehrhardt:2019}: The bins are optimized to maximize either AIC, BIC or the Gini coefficient (cf. Sec.~\nameref{sec:performance}) of a subsequent logistic regression model (using binned variables, not WoEs) on validation data (argument \code{criterion =}). In addition, also second order interactions can be considered (argument \code{interact = TRUE}). Note that this approach is comparatively intense in terms of computation time and does not take into acount for variable selection (cf. Sec.~\nameref{sec:scorecard}).  

Some packages do not provide their own implementations of an automatic binning but just interfaces to discretization functions within other packages: 
\code{Rprofet::BinProfet()} uses the function \code{greedy.bin()} of the package \CRANpkg{binr} \citep{binr}. The package \pkg{scoringTools} contains a bunch of functions (\code{chiM\_iter()}, \code{mdlp\_iter()}, \code{chi2\_iter()}, \code{echi2\_iter()}, \code{modchi2\_iter()} and \code{topdown\_iter()})
which provide interfaces to binning algorithms from the package \CRANpkg{discretization} \citep{discretization}. The \CRANpkg{dlookr} package \citep{dlookr} which is primary designed for exploratory data analysis has an implemented interface (\code{binning\_by()}) to \code{smbinning::smbinning()}.

\noindent {\bf Manipulation of the bins and assigning bins to data:} 

As it has been outlined before manual inspection and manipulation of the bins is considered as a substantial part of the scorecard development process. Two of the aforementioned packages provide functions to support this.  \code{scorecard::woebin()} allows to pass an argument \code{breaks\_list}: Each element corresoponds to a variable with manual binning and must be named like the corresponding variable. For numeric variables it must be a vector of break points and for factor variables it must be a character vector of the desired bins given by the merged factor levels, separated by \code{"\%,\%"} (cf. output from example 3 for variable purpose). In addition, a function \code{scorecard::woebin\_adj()} allows for an interactive adjustment of bins.
The package \pkg{smbinning} provides two functions \code{smbinning.custom()} and \code{smbinning.factor.custom()}. 

Manipulation of the bins should be based on an analysis of the binning results. For this purpose, most of the packages provide result tables on a variable level. The subsequent code example illustrates the step of an initial automatic binning as created by the package \pkg{scorecard}: 
%as well as the corresponding results table for the variable \code{purpose}:

\pagebreak

\begin{example}
### example 2: automatic binning
library(scorecard)
bins         <- woebin(train, y = "creditability", method = "tree")

# binning results for variable purpose
options(digits = 3)
bins$purpose[,c(2,3,4,5,6,7,8,10)]

# vizualize bins for variable purpose
woebin_plot(bins, x = "purpose", line_value = "woe")
\end{example} 
%$

The resulting table contains several key figures for each bin: the distribution (absolute and relative frequency of the samples given the level of the target variable), default rate and optionally the bin's WoE as well as the IV of the binned variable: 

{\footnotesize
\begin{example}
                                       bin count count_distr good bad badprob    woe total_iv
1:                    business%,%car (new)   227      0.3211  148  79   0.348  0.213    0.161
2:                              car (used)    77      0.1089   67  10   0.130 -1.061    0.161
3:         domestic appliances%,%education    44      0.0622   24  20   0.455  0.659    0.161
4:            furniture/equipment%,%others   137      0.1938   90  47   0.343  0.192    0.161
5: radio/television%,%repairs%,%retraining   222      0.3140  165  57   0.257 -0.222    0.161
\end{example}
}

In addition to summary tables many packages (\pkg{glmdisc}, \pkg{riskr}, \pkg{Rprofet}, \pkg{scorecard}, \pkg{smbinning}, \pkg{woeBinning}) provide a vizualization of the bins on a variable level. Fig.~\ref{figure:bins} (left) shows the binning as resulting from code in example 2 which is very similar for most packages. A mosaic plot of the bins which simultaneously visualizes default rates as well as size of the bins is offered by the package \pkg{glmdisc} (Fig.~\ref{figure:bins}, right) while the names of the bins after automatic binning are not self-explanatory. 

\begin{figure}[htbp]
  \centering
  \includegraphics[width = 14cm]{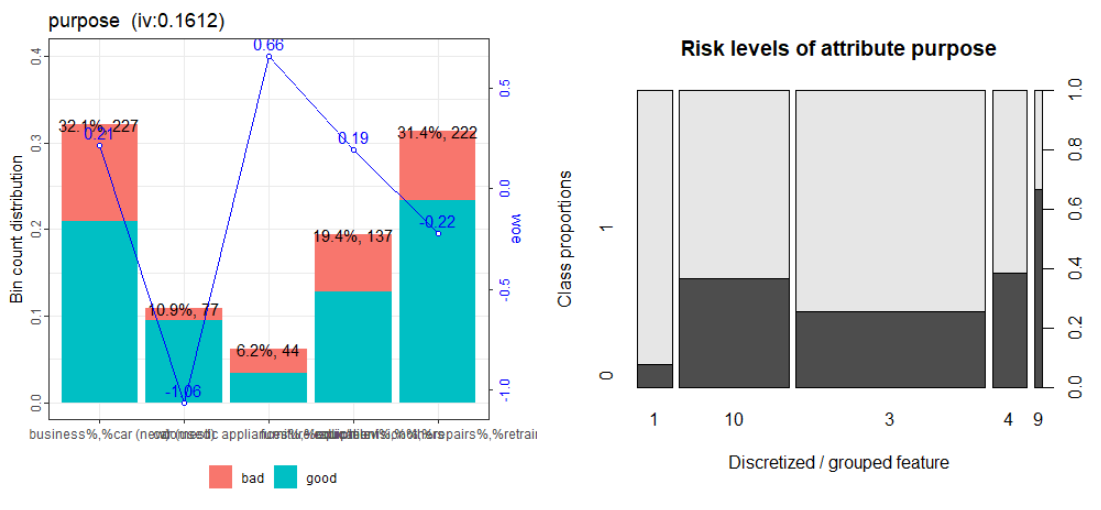}
  \caption{Visualization of the bins for the variable \code{purpose} as created by the package \pkg{scorecard} (left) and mosaicplot of the binning result by the package \pkg{glmdisc} (right).}
    \label{figure:bins}
\end{figure}

\noindent {\bf Applying bins to new data:} 

It has been emphasized that the bins as they are built on training data do constitute the first part of a scorecard model. 
For this reason it is necessary to store the results of the binning and to have functions to apply it to a data set. 

Most of the packages do provide this functionality such as \pkg{scorecard} (\code{woebin\_ply()}), \pkg{smbinning} (\code{smbinning.gen()} and \code{smbinning.factor.gen()}), \pkg{woeBinning} (\code{woe.binning.deploy()}), \pkg{woeR} (\code{apply\_woe()}\footnote{Supports only WoE predictions, not bins.}), \pkg{glmdisc} (\code{discretize()}) and \pkg{scorecardModelUtils} (\code{num\_to\_cat()}). Example 3 illustrates application of binning results to a data set. Via the \code{to = "bin"} argument either bins or WoEs can be assigned:   

\begin{example}
### example 3: apply binning to data
train_bins <- scorecard::woebin_ply(train, bins, to = "bin")
valid_bins <- scorecard::woebin_ply(valid, bins, to = "bin")
\end{example} 

For \code{ctree}-based binning (cf. above) a workaround using the \code{partykit::predict.party()} method for bin assignment can be obtained if the tree model is stored within the results object\footnote{An example code for the package \pkg{riskr} is given in snippet 2 of the supplementary code.}.   
More generally, binned variables can be created via the function \code{cut()} for numeric variables or by using lookup tables for factor variables \citep[cf. e.g.][p.23]{zumel:2014}\footnote{An example using a lookup table for the variable \code{purpose} is given in snippet 3 of the supplementary code.}.   
It is worth mentioning that several packges (\pkg{smbinning}, \pkg{woeR} and \pkg{riskr}) implement binning only on a single variable level but not simultaneously for several selected variables or all variables of a data frame\footnote{A code example of looping through all (numeric) variables for the package \pkg{smbinning} is given in snippet 4 of the supplementary code.}.

\noindent {\bf Binning of categorical variables:} 

For categorical variables, initially, each level can be considered as a separate bin but levels of similar default rate and/or meaning could be grouped together. As an additional challenge there is no natural order of the levels. For these reasons only some of the packages do offer an automatic binning of categorical variables. 
For example the package \pkg{smbinning} does not offer an automatic merging of levels for factor variables and its function \code{smbinning.factor()} only returns the figures similar to the table resulting from example 2 but each original level just corresponds to one bin. The bins can be manipulated afterwards via \code{smbinning.factor.custom()} and further be applied to new data via \code{smbinning.factor.gen()}. 
An automatic binning of categorical variables based on conditional inference trees is suppoerted by the packages \pkg{riskr} and \pkg{scorecard} (\code{method = "tree"}).
Additional merging strategies are provided by the packages \pkg{glmdisc} (as described above), \pkg{scorecard} (\code{method = "chimerge"}) and \pkg{woeBinning} (according to similar WoEs).  

Generally, merging of levels with similar default rate should only be done if level's frequency is large enough to result in a reliable default rate estimate on the sample. 
By using \pkg{woeBinning}'s \code{woe.binning()} function this can be ensured: Initial bins of a minimum size (\code{min.perc.total}) are created and smaller factor levels are initially bundled into a positive or negative 'miscellaneous' category according to the sign of the corresponding WoE which is desirable in order to prevent from overfitting. 
The package \pkg{scorecardModelUtils} offers a separate function \code{cat\_new\_class()} for this:  
All levels less frequent than specified by the argument \code{threshold} are merged together and 
a data frame with the resulting mapping table is stored in the output element \code{\$cat\_class\_new}\footnote{An example code for application of this mapping to new data is given in snippet 5 of the supplementary code. The names of the resulting new levels are the concatenated old levels, separated by commas. Note that the function can not deal with commas in the original level names: a new level \code{<NA>} will be assigned}. 

Similar to \pkg{woeBinning}'s  \code{woe.binning()} function \code{woebin()} of the \pkg{scorecard} package also merges adjacent levels of similar default rate for categorical variables (using \code{method = "chimerge"}).  
An important difference between both implementations consists in how the deal with the missing natural order of the levels and thus the notion of what 'adjacent' means: 
In contrast to \code{woe.binning()} where the levels are sorted according to their WoE before merging this is not the case for the function\code{woebin()} and levels are merged along their natural order which will often be alphabetical order\footnote{This can be easyily checked using the variable \code{purpose}, cf. e.g. snipplet 6 of the supplementary code.}. 
This might lead to an undesired binning and as an important conclusion an analyst should think about manually changing the level order for factor variables when working with the package \pkg{scorecard}\footnote{A code snippet for creating a \code{breaks\_list} (cf. above) from a binning result using the package \pkg{woeBinning} that can be imported for further use within the package \pkg{scorecard}, e.g. for manual manipulation of the bins is given by the function \code{woeBins2breakslist()} in snippet 7 of the supplementary code}.

\noindent {\bf Weights of Evidence:} 

Most of the above-mentioned packages do provide WoEs of the bins within their binning summary  tables. In order to use WoEs within the further modelling steps it in addtion needs for a functionality to assign the corresponding WoE value for each bin to the original (/or binned) variables as given by \code{scorecard::woebin\_ply()} (with argument \code{to = "woe"}), \code{woeBinning::woe.binning.deploy()} (with argument \code{add.woe.or.dum.var = "woe")} and (for single variables) \code{woeR::apply\_woe()}.

A quite general way of training, storing and assigning WoEs indepently of the package used for binning is given by the function \code{woe()} in the \pkg{klaR} package, probably the first and most comprhensive implementation of WoE computation in R. WoEs for binned variables are computed on the training data and stored in an \code{S3} object of class \code{woe} with a corresponding \code{predict.woe()} method that allows application to new data. Further, via an argument \code{ids} a subset of the variables can be selected for which WoEs are to be computed (default: all \code{factor} variables) as well as a real value \code{zeroadj} can be specified that is added to the frequency of bins with empty target levels for computation of $f()$ in eqn.~1 in order to prevent from WoEs resulting in $\pm \infty$. 
As a difference to other implementations it also allows to assign observation \code{weights} which can be necessary for \nameref{sec:ri}. The subsequent code shows its usage:

\pagebreak

\begin{example}
### example 4: computing and applying WoEs (based on example 3)
library(klaR)
# woe() requires variable type factor 
train_bins <- dplyr::mutate_if(train_bins, sapply(train_bins, is.character), as.factor)
valid_bins <- dplyr::mutate_if(valid_bins, sapply(valid_bins, is.character), as.factor)

# Compute WoEs on training data
woe_model <- woe(creditability ~ ., data = train_bins)
# ...woes for variable purpose
woe_model$woe$purpose_bin

# apply WoEs
train_woes <- data.frame(creditability = train_bins$creditability, 
                         woe_model$xnew)
valid_woes <- predict(woe_model, valid_bins)
\end{example}

\noindent {\bf Short benchmark experiment:}

The example data has been used to compare the performance of the different available packages for automatic binning. For reasons explained above binning of categorical variables requires expert's knowledge on the meaning of the levels thus the benchmark is restricted to a comparison for the seven numeric variables in the data set. Note that four of these variables do contain small numbers of distinct numeric values such as e.g. the number of credits (cf. $2^{nd}$ column of table~1). Therefore the remaining three variables \code{age}, \code{amount} and \code{duration} are the most interesting ones. 
Further note that (although it is by far the most popular data set used in literature) for reasons of its size and the balance of the target levels the German credit data might not be representative for typical credit scorecard developments \citep{szepannek:2017wias}. For this reason the results should not be overemphasized but rather give an idea on differences in performance of the various implementations. 

\begin{table}[ht]
\centering
\footnotesize
\begin{tabular}{rrrrrrrrrrr}
  \toprule
 & unique & sc & woeB & woeB.T & glmdisc & scMU & Rprof & smb & woeR & riskr \\ 
  \midrule
  Avg. \# bins   	&  & 6.33 & 4.33 & 6 & 2.33 & 3.67 & 11 & 2.67 & 3.33 & 2.67 \\     \midrule
  age 		& 52 	& 7 & 4 & 7 & 4 & 3 & 9 & 2 & 2 & 2 \\ 
  amount 	& 663 	& 7 & 4 & 6 & 1 & 3 & 11 & 3 & 3 & 3 \\ 
  duration 	& 32 	& 5 & 5 & 5 & 2 & 5 & 13 & 3 & 5 & 3 \\ 
  instRate 	& 4 	& 4 & 4 & 4 & 2 & 4 & 4 & 1 & 3 & 1 \\ 
  numCredits & 4 	& 2 & 3 & 3 & 2 & 2 & 3 & 1 & 2 & 1 \\ 
  numLiable  & 2 	& 2 & 3 & 3 & 2 & 1 & 2 & 1 & 1 & 1 \\ 
  residence & 4 	& 4 & 4 & 4 & 3 & 2 & 4 & 1 & 2 & 1 \\ 
  \bottomrule
\end{tabular}
\label{tab:nbins}
\caption{Number of bins after automatic binning. Abbreviations of package names: sc = \pkg{scorecard}, woeB = \pkg{woeBinning} using \code{woe.binning()}, woeB.T = \pkg{woeBinning} using \code{woe.tree.binning()}, scMU = \pkg{scorecardModelUtils}, Rprof = \pkg{Rprofet} and smb = \pkg{smbinning}.} 
\end{table}

Table~1 shows the number of bins as returned by the different packages. The first row summarizes the average number of bins for the three variables\code{age}, \code{amount} and \code{duration}: The package \pkg{Rprofet} (which interfaces to \code{binr::bins.greedy()}, cf. above) returns the largest number of bins. The number of bins as returned by the tree-based binning via \pkg{smbinning} and \pkg{riskr} as well as \pkg{glmdisc} are comparatively small.

\begin{table}[ht]
\centering
\footnotesize
\begin{tabular}{rlrrrrrrrrr}
  \toprule
 			& Package 	& LCL 	& sc 	& woeB 	& woeB.T 	& glmdisc 	& scMU 	& smb 	& woeR 	& riskr \\ 
  \midrule
  age 		& woeB.T 	& 0.078 & 0.179 & 0.169 & 0.222 	& 0.177 	& 0.200 & 0.187 & 0.061 & 0.187 \\ 
  amount 	& sc 		& 0.116 & 0.251 & 0.179 & 0.227 	& 0.000 	& 0.196 & 0.219 & 0.205 & 0.219 \\ 
  duration 	& scMU,woeR 		& 0.170 & 0.297 & 0.259 & 0.264 	& 0.155 	& 0.299 & 0.248 & 0.299 & 0.248 \\ 
  instRate 	& woeR 		& 0.000 & 0.108 & 0.103 & 0.103 	& 0.104 	& 0.108 & 0.000 & 0.111 & 0.000 \\ 
  numCredits &  		& 0.000 & 0.068 & 0.068 & 0.068 	& -0.010 	& 0.068 & 0.000 & -0.040 & 0.000 \\ 
  numLiable &  			& 0.000 & 0.006 & 0.006 & 0.006 	& 0.006 	& 0.000 & 0.000 & 0.000 & 0.000 \\ 
  residence & scMU 		& 0.000 & 0.006 & 0.017 & 0.017 	& 0.002 	& 0.029 & 0.000 & -0.045 & 0.000 \\ 
  \bottomrule
\end{tabular}
\label{tab:validation_ginis_binning}
\caption{Gini coefficient of WoE transformed variables on validation data.} 
\end{table}

Table~2 lists the performance of the different binning algorithms. In order to prevent from analyzing the overfitting of the training data (as it would be obtained by increasing the number of bins) the validation data is used for comparison (cf. example 1). 
In order to ensure a fair comparison of all packages the perfomance is computed using the same methodology: First, WoEs are assigned to the validation data using the package \pkg{klaR}. Afterwards, univariate Gini coefficients (as one of the most commonly used performance measures for performance evaluation of credit scoring models, cf. Sec.~\nameref{sec:performance}) of the WoE variables are computed using the package \CRANpkg{pROC} \citep{pROC}. 
Note that some of the introduced functions for automatic binning allow for a certain degree of \emph{hyperparameterization} which can be used to improve the binning results but as the scope of automatic binning does not consist in providing a highly tuned perfect model but rather a solid basis for a subsequent manual bin adjustment all results in the experiment are computed using default parameterization. 
Further note that for the package \pkg{Rprofet} no validation performance is available as there exists no \code{predict()} method. For the packages \pkg{riskr} the workaround has been used as described above to assign bins to validation data\footnote{The corresponding code is given in the file benchmark.R}. Concerning the results it also has to be mentioned that the package \pkg{glmdisc} optimizes bins w.r.t. subsequent logistic regression based on dummy variables on the bins which of course further takes into account the multivariate dependencies between the variables and not just discriminative power of the single variables\footnote{Note that the call of \code{glmdisc()} ran in an internal error (\code{Error in t * y\_p : non-conformable arrays}) for more than 10 iterations. For this reason the number of iterations has been reduced to 10 which is much smaller than the default of 1000 iterations and the reported gini coefficient does still strongly vary among subsequent iterations. For larger numbers of iterations better results might have been possible.}. 

The first column (LCL) of the results contains a $95\%$ lower confidence level of the best binning for each variable using boostrapping \citep{robin:2011}: Only in two cases (package \pkg{woeR} for the variable \code{age} and \pkg{glmdisc} for the variable \code{amount} and \code{duration}) the bins are significantly worse (below LCL) than the best observed result and in summary none of the packages clearly dominates the others. On the first glance the choice of the algorithm does not seem to be crucial. It migth be worth trying different algorithms and comparing their resuls.

\noindent {\bf Summary of packages for variable binning:} 

Table~3 summarizes the functionalities of the different packages for the purpose of binning and WoE assignment as presented above.
 
\begin{table}[ht]
\centering
\footnotesize
\begin{tabular}{rlllllllll}
  \toprule
 & sc & smb & woeB & woeR & riskr & glmdisc & scMU & Rprof & klaR \\ 
  \midrule
automatic binning of numerics & x & x & x & x & x & x & x & x & o \\ 
  automatic binning of factors & x & o & x & o & x & x & o & o & o \\ 
  store and predict numerics & x & x & x & x & (*) & x & x & o & o \\ 
  store and predict factors & x & x & x & o & (*) & x & o & o & o \\ 
  supports bin prediction & x & x & x & o & (*) & x & x & o & o \\ 
  supports WoE prediction & x & o & x & x & o & o & o & o & x \\ 
  \midrule
  summary table & x & x & x & x & x & o & x & o & x \\ 
  plot & x & x & x & o & x & x & o & x & x \\ 
  manual modification & x & x & o & o & o & o & o & x & o \\ 
  \midrule
  multiple variables & x & o & x & o & o & x & x & x & x \\ 
  supported target levels & x & o & x & x & o & o & o & o & x \\ 
  adjust WoEs & x & o & x & o & o & r & x &  & x \\ 
  NAs & x & x & x & x & o & (**) & x/o & x &  \\ 
  new levels & o & o & x &  & o & (**) &  &  &  \\ 
  level order irrelevant & o &  & x &  & x & x &  &  &  \\ 
  min. level size & o &  & x &  & o & o & (***) & o &  \\ 
   \bottomrule
\end{tabular}
\label{Tab:summary_binning}
\caption{Summary of the functionalities for binning and WoEs provided by the different packages where 'x' denotes available and 'o' not available. An empty field means that this is not relevant w.r.t. the scope of the package. (*): workaround available (cf. above); (**) always bin 1 assigned; 
(***) additional function cat\_to\_new() merges levels smaller than threshold (cf. above).} 
\end{table}

In summary, the package \pkg{woeBinning} offers a quite comprehensive toolbox with many desirable implemented functionalities but unfortunately no manual modification of the results from automatic binning is supported. For the latter the \pkg{scorecard} package can be used which in contrast has to be used with care for factor variables as its automatic binning of categorical variables suffers from dependence on the natural order of the factor levels. As a remedy, a function has been suggested in the supplementary code (cf. footnote 9) to import the results of \pkg{woeBinning}'s automatic binning into result objects from the \pkg{scorecard} package for further processing.

\section{Preselection of variables}\label{sec:preselection}
As it has already been outlined above a major aspect of credit risk scorecard development is to allow for the integration of expert domain knowledge at different stages of the modelling process. 
For the puropose of variable selection in Statistics traditionally criteria such as AIC or BIC are used that trade-off model's performance vs. the number of free model parameters (cf. Sec.~\nameref{sec:scorecard}). For scorecard modelling, in addition typically a {\em variable preselection} is done which allows for a plausibility check by analysts and experts. Apart from plausibility checks several analyses are carried out at this stage typically consisting of:
\begin{itemize} 
	\item information values of single variables, 
	\item population stability analyses of single variables on recent out-of-time data as well as
	\item correlation analyses between variables. 
\end{itemize}

\noindent {\bf Information value:} 

Variables with small discriminatory power in terms of their IV (cf. Sec. \nameref{sec:binning}) are candidates to be considered to be removed from the further development process. While the interpretation ''small'' in the context of IV slightly varies depending on who is asked an example is given in \citet{siddiqi:2006} by $IV < 0.02$. 
As an important remark and in contrast to common practice in credit scorecard modelling 
not just the IV of a variable should be taken into account but rather how much {\em different} information a variable will contribute to a scorecard model that is not already included in other variables. For this reason, IVs should be analyzed together with correlations (cf. this Section below). 
If not just validation data but also an independent test data set is available a comparison of the IV on training and validation data can be used to check for overfitting of the binning.

Table~4 lists packages that provide functions to compute information values of binned variables. As usual these packages differ by the type of the target variable that is required: some of them do allow for factors others require binary numerics that take the values 0 and 1. An important difference consists in whether (and how) they do WoE adjustment in case of bins where one of the classes is empty. In \pkg{creditR} no adjustment is done and the resulting IV becomes $\infty$. Some packages (\pkg{Information}, \pkg{InformationValue} and \pkg{smbinning}) do return a value different from $\infty$ but from the documentation it is not clear how it is computed. For the packages \pkg{scorecard} and \pkg{scorecardModelUtils} the adjustment is known and for the package \pkg{klaR} the adjustment can be specified in an argument. Note that depending on the adjustment the resulting IVs of the affected variables may differ strongly. 

\begin{table}[ht]
\centering
\footnotesize
\begin{tabular}{llccc}
  \toprule
  Package           		& Function                    	& Target type    & multiple variables & WoE adjustment\\ 
  \midrule
  \pkg{creditR} 			& \code{IV.calc.data()} 		& both, levels 0/1 	& yes & no \\ 
  \pkg{Information} 		& \code{create\_infotables()}   & numeric 0/1	 	& yes & yes \\ 
  \pkg{InformationValue} 	& \code{IV()} 					& numeric 0/1 		& no  & yes \\ 
  \pkg{klaR} 				& \code{woe()} 					& factor	 		& yes & argument \\ 
  \pkg{riskr} 				& \code{pred\_ranking()} 		& numeric 0/1 		& yes & no \\ 
  \pkg{scorecard} 			& \code{iv()} 					& both	 			& yes & 0.99 \\ 
  \pkg{scorecardModelUtils} & \code{iv\_table()} 			& numeric 0/1 		& yes & 0.5 \\ 
  \pkg{smbinning} 			& \code{smbinning.sumiv()} 		& numeric 0/1		& yes & yes \\ 
  \bottomrule
\end{tabular}
\label{Tab:summary_IV}
\caption{Packages and functions for computation of IVs.} 
\end{table}

Example 5 shows how IVs can be computed using the package \pkg{klaR} with zero adjustment (which in fact is not necessary here.) The function \code{woe()} (cf. example 4) automatically returns IVs for all factor variables.

\begin{example}
### example 5: computing IVs  (based on example 4)
library(klaR)
woe_model <- woe(creditability ~ ., data = train_bins, zeroadj = 0.5)
# ...the IVs are automatically computed and can be assessed via: 
woe_model$IV
\end{example}  
%$

The package \pkg{creditR} also offers a function \code{IV\_elimination()} which allows to set an \code{iv\_threshold} and returns a data set with a subset of variables with IV above threshold for the training data. Similarly, the package \pkg{scorecardModelUtils} offers a function \code{iv\_filter()} that returns a list of variable names that pass (/fail) a prespecified threshold.

Beyond computation of IVs the package \pkg{creditR} can be used to compute Gini coefficients for simple logistic regression models on each single variable via the function \code{Gini.univariate.data()} and just as for IVs this can be used for variable subset preselection (\code{Gini\_elimination()}). The function \code{pred\_ranking()} from the package \pkg{riskr} returns a summary table containing IV as well as the values of the univariate AUC and KS statistic and an interpretation.

\noindent {\bf Population stability analysis:} 

In order to take into account for sample selection bias that results from a customer portfolio shift (e.g. due to new products or marketing activities) the stability of the distribution of the variable's bins over time is considered. For this purpose typically the {\em population stability index (PSI)} is computed between the (historical) development sample data and a more recent {\em out-of-time (OOT) sample} (where typically performance information is not yet available). Basically the PSI is just the IV (cf. eqn. (2)). While the IV compares two data sets given by the development sample which is split according to the levels of the target variable ($y=1$ vs. $y=0$) the PSI compares the entire development sample ($y \in \{0,1\}$) with an entire out-of-time sample. A large PSI indicates a change in the population w.r.t. the bins. A small PSI close to 0 indicates a stable population and \citep[again refering to][]{siddiqi:2006}  $PSI < 0.1$ can be interpreted as stable while a $PSI > 0.25$ is an indicator for a population shift. Of course a decision of inclusion or removal of variables from the development sample should take into account for both population stability as well as the discriminatory power (i.e. IV) of a variable. 
With reference to the analogy for PSI and IV the formerly presented functions of IV calculation can also be used for population stability analysis. 
The function  \code{SSI.calc.data()} from the package \pkg{creditR} returns a data frame of PSIs for all variables. 
The corresponding code (here for a computation of PSIs between training and validation  -- not OOT -- set) is given in example 6.   

\begin{example}
### example 6: population stability analysis for all variables
library(creditR)
SSI.calc.data(train_bins, valid_bins, "creditability")
\end{example} 

The function \code{riskr::psi()} calculates the PSI for single variables and also provides a more detailed table on the bin-specific differences (cf. example 7 for the variable \code{purpose}). 

\begin{example}
### example 7: population stability analysis for a single variable
library(riskr)
# PSI for binned variable purpose (based on example 3)
psi(train_bins$purpose_bin, valid_bins$purpose_bin)
\end{example} 

The results table for the variable \code{purpose} does contain the absolute and relative distribution of the bins (for reasons of space two columns with the absolute frequencies have been discarded from the output). The PSI of the variable as given by the \code{value} element of the output corresponds to the sum of the column \code{index}:

{\scriptsize
\begin{example} 
$value
[1] 0.00792

$label
[1] "Insignificant change"

$table
# A tibble: 5 x 9
  class                                   act_percent new_percent diff_percent coefficient     woe     index
  <chr>                                         <dbl>       <dbl>        <dbl>       <dbl>   <dbl>     <dbl>
1 business%,%car (new)                         0.321       0.355      0.0339         1.11   0.100  0.00340  
2 car (used)                                   0.109       0.0887    -0.0202         0.815 -0.205  0.00413  
3 domestic appliances%,%education              0.0622      0.0614    -0.000801       0.987 -0.0130 0.0000104
4 furniture/equipment%,%others                 0.194       0.191     -0.00265        0.986 -0.0138 0.0000365
5 radio/television%,%repairs%,%retraining      0.314       0.304     -0.0102         0.967 -0.0332 0.000340 
\end{example} 
%$
}

Alternatively the package \pkg{smbinning} comes along with a function \code{smbinning.psi(df, y, x)} which requires both development and OOT sample to be in one data set (\code{df}) and an variable \code{y} that indicates the data set where an observations originates from. The packages \pkg{creditR} and \pkg{scorecard} offer functions which can be used for an OOT stability analysis of the final score %which will be described in 
(cf. Sec.~\nameref{sec:scorecard}).

\noindent {\bf Correlation analysis:} 

\begin{figure}[htbp]
  \centering
  \includegraphics[width = 14cm]{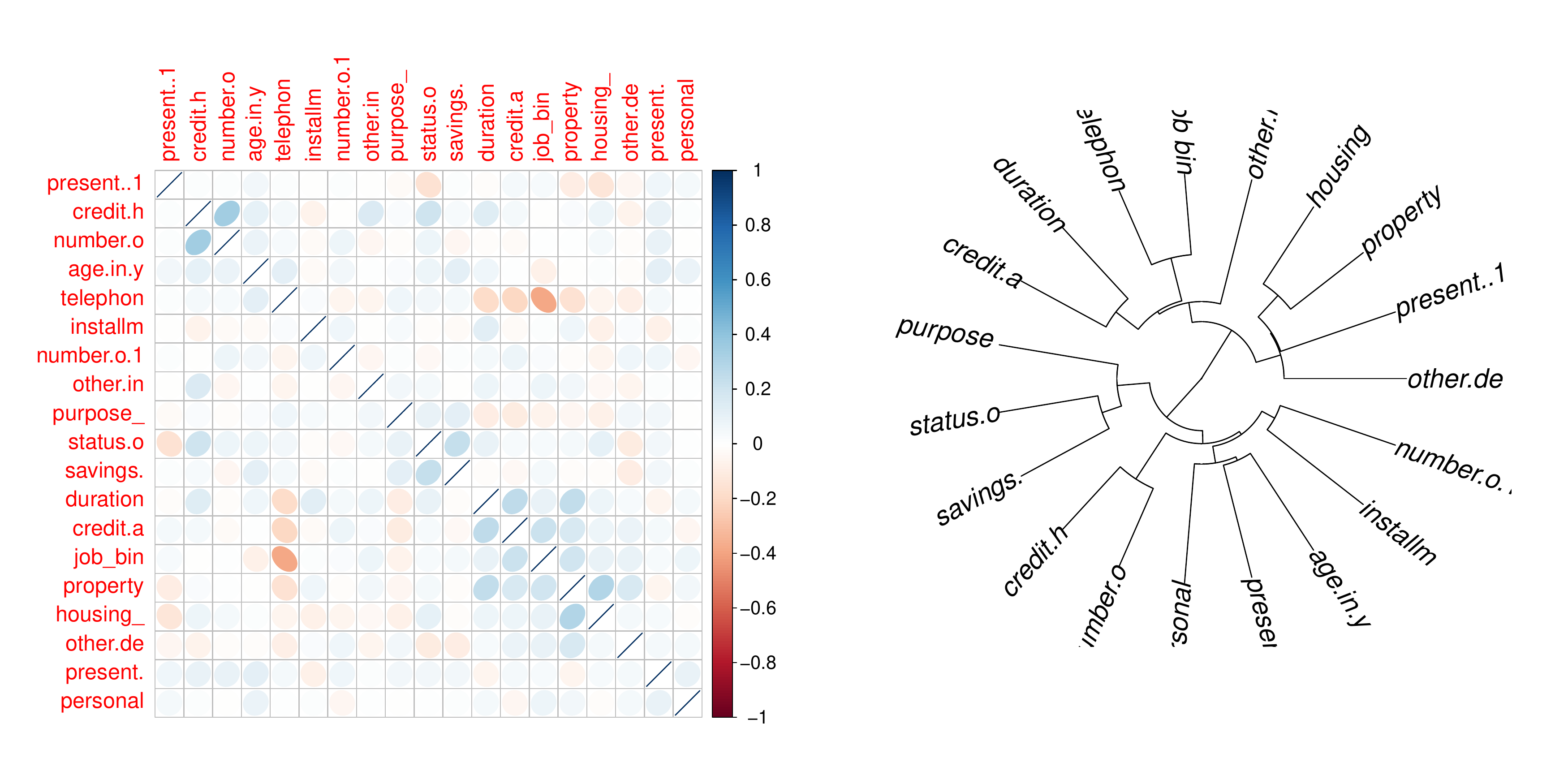}
  \caption{Reordered correlation matrix using the package \pkg{corrplot} (left) and phylogenetic tree of the clustered variables using the packages \pkg{ClustOfVar} and \pkg{ape} (right).}
    \label{figure:cor}
\end{figure}

In order to avoid variability of the estimates of a regression model its regressors should be of low correlation \citep[cf. e.g.][ch. 3,4]{hastie:2009}. 
As per construction WoE transformed variables are linear in the logit of the target variable a natural approach consists in analyzing correlations between these variables.
For this purpose the \pkg{caret} package (\citealp{caret, kuhn:2008}) offers a function \code{findCorrelation()} that automatically identifies among any two variables of strong correlation the one that has larger average (absolute) correlation to all other variables. 
A major advantage of performing correlation analysis in advance for variable preselection is that it can be used as another way to integrate expert's experience into the modelling: Among variable clusters of high correlations experts can choose which of these variables should be used or discarded for further modelling. There are some packages not originally designed for credit scorecard modelling which offer functions that can be used for this purpose: 
The package  \CRANpkg{corrplot} \citep{corrplot} offers a function to visualize the correlation matrix and resort it such that groups of correlated variables are next to each other (cf. Figure~\ref{figure:cor}, left). An alternative visualization is given by a {\em phylogenetic tree} of the clustered variables using the package \CRANpkg{ape} (\citealp{ape, paradis:2018}) where the variable clustering is obtained using the package \CRANpkg{ClustOfVar} (\citealp{ClustOfVar, chavent:2012}, cf. Figure~\ref{figure:cor}, right). The code for creation of both plots is given in the following example (note that the choice of the \code{hclust.method = "complete"} in the left plot guarantees for a minimum correlation among all variables in a cluster but all correlations on the training data are below $0.35$ in this example): 
            
\begin{example}
### example 8: visualizing correlations (based on example 4)
# reordered correlation matrix
library(corrplot)
# crop redundant prefixes from variable names for plot
X <- train_woes
names(X) <- substr(names(X), 5, 12)
cmat <- cor(X[,-(1:2)])
corrplot(cmat, order = "hclust", method = "ellipse", hclust.method = "complete")

# phylogenetic tree
library(ClustOfVar)
library(ape)
vctree  <- hclustvar(X.quanti = X[,-(1:2)]) 
plot(as.phylo(vctree), type = "fan")
\end{example} 

The package \CRANpkg{clustVarLV} (\citealp{ClustVarLV, vigneau:2015}) offers variable clustering such that the correlation between each variable and the first latent principal component of its variable cluster is maximized. The number of clusters \code{K} has to be pre-specified. As it can be seen in the output from example 9 (only cluster 1 is shown) for each variable the correlation to the cluster's latent component as well as the correlation to the 'closest' next cluster are shown.

\begin{example}
### example 9: variable clustering using ClustVarLV
library(ClustVarLV)
clverg <- CLV(train_woes[,-(1:2)], method = 1)
plot(clverg)
summary(clverg, K = 3)  
\end{example} 

\begin{example}
### Output of example 9 for variable cluster no.1:
                                            cor in group  |cor|next group
woe_savings.account.and.bonds_bin                   0.73             0.05
woe_status.of.existing.checking.account_bin         0.72             0.17
woe_purpose_bin                                     0.51             0.11
\end{example} 

Among the aforementioned packages dedicated to credit scoring, \pkg{creditR} contains a function \code{variable.clustering()} that performs \CRANpkg{cluster}'s \code{pam} \citep{cluster} on the transposed data for variable clustering. The (sparsely documented) function 
\code{correlation.cluster()} %data, output, "variable", "Group")} 
can be used to compute average correlations between the variables of each cluster. \footnote{Its argument \code{data} denotes the training data, \code{output} is a data frame with two variables specifiying the variable names of the training data (\code{character}) and the corresponding cluster index, as given e.g. by the result from \code{variable.clustering()}. Finally, its arguments \code{variables} and \code{clusters} denote the names of these two variables in the data frame from the \code{output} argument  where the clustering results are stored.}

The package \pkg{Rprofet} provides two functions \code{WOEClust\_hclust()} and  \code{WOEClust\_kmeans()} that do perform \code{stats::hclust()} on the transformed data or \code{ClustOfVar::kmeansvar()} and return a data frame with variable names and cluster index together with the IV of the variable which may help to select variables from the clusters. Unfortunately, they are only designed to work with output from the package's function \code{WOEProfet()} and require a list of a specific structure as input argument. 

Notably, the package \pkg{scorecardModelUtils} offers an alternative for an automatic variable preselection based on {\em Cramer's V} using the function \code{cv\_filter()}: Among two (categorical) variables of $V >$ \code{threshold} the one with lower IV is automatically removed (cf. example 10). Finally, two functions \code{iv\_filter()} and \code{vif\_filter()} can be used for variable preselection based on IVs only (w/o taking into account for correlations between the explanatory variables) and based on variance inflation (cf. also Sec. \nameref{sec:scorecard}).  

\begin{example}
### example 10: Cramer's V based variable selection (based on example 3)
library(scorecardModelUtils)
# package requires 0/1 target:
train_bins2 <- train_bins
train_bins2$creditability <- as.integer(train_bins2$creditability == "good")

# first data frames of IVs and Cramer's V have to be computed
ivtable   <- iv_table(train_bins2, "creditability", cat_var_name = names(train_bins2)[-1])
cvtable   <- cv_table(train_bins2, names(train_bins2)[-1])
selection <- cv_filter(cvtable$cv_val_tab, ivtable$iv_table, threshold = 0.3)
selection 
\end{example}

\noindent {\bf Further useful functions to support variable preselection:} 

The package \pkg{scorecard} contains a function \code{var\_filter()} that performs an automatic variable selection based on IV and further allows to specify a maximum percentage of missing or identical values within a variable but not takes into account for correlations among the predictor variables.

The package \pkg{creditR} has two functions to identify variables with \emph{missing values} (\code{na\_checker()}) and compute the percentage of variables with missing values (\code{missing\_ratio()}). For imputation of numeric variables in a data set with mean or median values a function \code{na\_filler\_contvar()} is available. Of course, this has to be handled with care as mean or median value will typically not be the same on training and validation data. The package \CRANpkg{mlr} (\citealp{mlr, bischl:2016mlr}) offers imputation that can be applied on new data.

For an assignment of explicit values to missings the package \pkg{scorecardModelUtils} also provides a function \code{missing\_val()}. This can be either a function such as \code{"mean"}, \code{"median"} or \code{"mode"} or an explicit value such as -99999 which can be meaningful before binning in order to assign missing values to a separate bin. Similarly, for categorical variables the assignment of a specific level such as \code{"missing\_value"} can be meaningful. A function \code{missing\_elimination()} removes all variables with a percentage above \code{missing\_ratio\_threshold} from training (but not from validation) data.

The package \pkg{riskr} provides two functions \code{select\_categorical()} and \code{select\_numeric()} to select all (non-/) numeric variables of a data frame. 

A univariate summary of all variables is given by the function \code{univariate()} of the 
\pkg{scorecardModelUtils} package. A summary for numeric variables can be computed using the function \code{ez\_summ\_num()} from the package \pkg{riskr}. A general oveview of packages explicitly designed for exploratory data analysis that do provide further functionalities is given in \citet{staniak:2019}. The package \pkg{scorecard} contains two functions \code{one\_hot()} and \code{var\_scale()} for one-hot-encoding and standardization of numeric variables.

%\begin{itemize} 	
%	\item \code{creditR::variable.clustering.gini}? 
%\end{itemize}

\section{Scorecard modelling}\label{sec:scorecard}

Traditionally, credit risk scorecards are modelled by logistic regression \citep[cf. e.g.][]{finlay:2012} which is in R done via \code{glm()} (with \code{family = binomial}). 

\noindent {\bf Variable selection:} 

In addition to the manual variable preselection as described in the former Section typically a subsequent variable selection is performed which can be done by the \code{step()} function. Common criteria for variable selection are AIC (\code{k = 2}) or BIC (\code{k = log(nrow(data))}). Example 11 gives an example for BIC based variable selection.

\pagebreak

\begin{example}
### example 11: BIC variable selection (based on example 4)
# column 2 (variable foreign.worker_bin) excluded as binned variable has only one level  
null <- glm(creditability ~ 1, data = train_woes[,-2], family = binomial)
full <- glm(creditability ~ ., data = train_woes[,-2], family = binomial)
bicglm <- step(null, scope=formula(full), direction="both", k=log(nrow(train_woes)))
\end{example}

Note that an initial model (here: \code{null}) and the scope for the search have to be specified. This offers another possibility for expert knowledge integration: After each step the criteria of all candidates are reported and can be used to decide among several variable candidates of similar performance for the one which is most appropriate from a business point of view. The corresponding variable can be manually added to the formula of a new initial model in a subsequent variable selection step.

The function \code{smbinning.logitrank()} of package \pkg{smbinning} runs all possible combinations of a specified set of variables, ranks them according to AIC and returns the corresponding model formulas in the result data frame. Depending on the size of the preselected set of variables (cf. Sec.~\nameref{sec:preselection}) this can be time consuming.

As an alternative to AIC and BIC \citet{scallan:2011} presents how variables can be selected in line with the concept of information values (cf. Sec. \nameref{sec:binning}) using so-called \emph{marginal information values} but currently none of the presented packages offers an implementation of this strategy. 

It is also common to consider the \emph{variance inflation factor} of the explanatory variables of a final model given by:   
\begin{equation}
	VIF(X_i) = \frac{1}{1 - R^2_i}
\end{equation}
where $R^2_i$ is the $R^2$ of a linear regression model with $X_i$ as dependent variable and all other explanatory variables except $X_i$ as regressors. Large values of $VIF(X_i)$ denote that this variable can be explained by the other regessors and are an indication of multicollinearity. Both the package \CRANpkg{car} (\citealp{car, fox:2019}) as well as the package \pkg{scorecard} offer a function \code{vif()} that can be used for this purpose as well as the function \code{vif.calc()} from the package \pkg{creditR} (cf. example 12). 

\begin{example}
### example 12: VIF (based on example 11)
car::vif(bicglm)
scorecard::vif(bicglm)
creditR::vif.calc(bicglm)
\end{example}

A question might not only consist in selecting variables and fitting a model but also in \emph{segmentation} i.e. whether one single model is sufficient or rather several separate models should be used. For this purpose the package \pkg{glmtree} offers a function \code{glmtree()} that computes a potential segmentation scheme according to a tree of recursive binary splits where each leaf of the tree consists in a logistic regression model. The resulting segmentation optimizes AIC, BIC or alternatively the likelihood or the gini coefficient on validation data. Note that this optimization does not take into account for variable selection as described above.

\noindent {\bf From logistic regression models to scorecards:} 

From the coefficients of the logistic regression model the historical shape of a \emph{scorecard} is obtained by assigning the corresponding effect (aka \emph{points}) to each bin (such that the score of a customer is the sum over all applicable bins and can easily calculated by hand). Typically, the effects are scaled in order to obtain some predefined \emph{points to double the odds} \citep[pdo, cf. e.g.][]{siddiqi:2006} and rounded to integers.

The package \pkg{scorecard} offers a function \code{scorecard()} that translates a \code{glm} object into scorecard points as described above and in addition returns key figures such as frequencies, default rates and WoE for all bins. A function \code{scorecard\_ply()} is available that can be used to assign scores to new data.
In addition to the \code{glm} object the \code{bins} as created by \pkg{scorecard}'s \code{woebin()} (cf. Sec.~\nameref{sec:binning}) have to be passed as an input argument. 
Further arguments do specify the (\code{pdo}) as well as a fixed number of points \code{points0} that corresponds to odds of \code{odds0} 
and whether the scorecard should contain an intercept or whether the intercept should be redistributed to all variables (\code{basepoints\_eq0}). The function requires WoEs (not just the binned \code{factor}s) and the variable names in the \code{coef(glm)} to match the convention of variable renaming as it is done by \pkg{scorecard}'s \code{woebin\_ply()} function (i.e. a postfix \code{\_woe})\footnote{A remedy how it can be used in combination with WoE assignment using the package \pkg{klaR} as shown in example 4 is given in snippet 9 of the supplementary code.}. 

Alternatively, a function \code{scorecard2()} is available which directly computes a scorecard based on \code{bins} and a data frame of the original variables. Here, in addition the name of the target variable (\code{y}) and a named vector (\code{x}) of the desired input variables have to be passed\footnote{Snippet 10 of the supplementary code illustrates how the vector \code{x} of the names of the input variables in the original data frame can be  extracted from the \code{bicglm} model after variable selection from example 11.}. Examples 13 and 14 illustrate the usage of \code{scorecard2()} and its application to new data (here represented by the validation set) as well as its output for the variable \code{duration.in.month}:    

\begin{example}
### example 13: calculation of scores (based on example 2)
sc2    <- scorecard2(bins, train, y = 'creditability', x = names(train)[1:19])
sc2       # note: variable 20 (foreign.worker) not used (cf. also example 11)
train_scored <- scorecard_ply(train, sc2, only_total_score = FALSE)
valid_scored <- scorecard_ply(valid, sc2, only_total_score = FALSE)

\end{example}

\begin{example}
### example 14: output from example 13 for the variable 'duration.in.month' 
> sc2$duration.in.month[,c(1,2,4,5,6,7,8,13)]
            variable       bin count_distr good bad   badprob        woe points
1: duration.in.month  [-Inf,8)  0.08062235   51   6 0.1052632 -1.2988228     65
2: duration.in.month    [8,16)  0.35785007  194  59 0.2332016 -0.3490774     18
3: duration.in.month   [16,34)  0.37906648  179  89 0.3320896  0.1424939     -7
4: duration.in.month   [34,44)  0.10466761   44  30 0.4054054  0.4582511    -23
5: duration.in.month [44, Inf)  0.07779349   26  29 0.5272727  0.9504426    -48
\end{example}
%$

In addition the package further contains a function \code{report()} which takes the data, the (original) names of all variables in the final scorecard model and a breaks list (cf. Sec.~\nameref{sec:binning} which can be obtained from the bins) as input arguments and generates an excel report summary of the scorecard model. Different sheets are reported with information and figures on the data, model, scorecard  points, model performance, as well as the binning figures for all variables of the model which can be used for model development documentation in practice.   

In order to translate a glm based on \code{factor} variables (bins instead of WoEs) into scorecard points the package \pkg{scorecardModelUtils} provides a function \code{scalling()}. Its output can be used to predict scores for new data by a function \code{scoring()} (cf. example 15):

\begin{example}
### example 15: scorecard points for model based on bins, not WoEs (based on example 4)
library(scorecardModelUtils)
# create glm using factor variables -- foreign worker excluded (cf. above) 
full_bins <- glm(creditability ~ ., data = train_bins[,-21], family = binomial)
# calculate scorecard points from effects
sc3 <- scalling(train_bins, "creditability", model = full_bins, point = 15, factor = 2)	
sc3

# apply scorecard to new data
scoring(valid_bins, target = "creditability", sc3)
\end{example}

Another implementation of calculating scorecard points from a glm object based on bins and not WoEs is given by the function \code{smbinning.scaling()} which comes along with a predict function \code{smbinning.scoring.gen()} that can be used to score new observations but requires that the binned variables have been generated with \code{smbinning.gen()} or \code{smbinning.factor.gen()} (cf. Sec. \nameref{sec:binning}). A function \code{smbinning.scoring.sql()} is available that transforms the resulting scorecard into SQL code.   

The package \pkg{Rprofet} also contains a function \code{ScorecardProfet()} for this purpose 
%which calculates a glm with corresponding scorecard points 
but only based on binning and WoEs as calculated by functions from the package itself (cf. Sec.~\nameref{sec:binning}) and no function is available for application of the scorecard points to new data. The function \code{scaled.score()} of the package \pkg{creditR} transforms posterior default probabilities into scores where any \code{increase} points double the odds (of non-default) and odds of  \code{increase} correspond to \code{ceiling\_score} points. In addition, the package \pkg{creditR} offers a function that can be used to \emph{re-calibrate} an existing glm on calibration data. A simple logistic regression is fit on the \code{calibration\_data} with only one single input variable: the predicted log odds by the current model.

\section{Performance evaluation}\label{sec:performance}
In credit scoring model performance evaluation is used not only for model selection but also for third-party assessments of an existing model by auditors or regulators and in order to drive future management decisions whether an existing model should be kept in place or whether it should be replaced by a new one. Note that as opposed to common practise in machine learning hyperparameter tuning typically there is no separate validation data used for model selection \citep[cf. e.g.][]{bischl:2012} but in credit scorecard modelling the validation data serves for independent model validation (corresponding to test data in frameworks such as \pkg{mlr3}). While this is less critical in case of simple models such as logistic regression it should still be kept in mind, especially if the model is benchmarked against more flexible machine learning models like support vector machines, random forests or gradient boosting \citep[cf. e.g.][]{hastie:2009}.

\noindent {\bf Discrimination:} 

The two most popular performance metrics for credit scorecards are the \emph{Gini coefficient} $Gini = 2 (AUC - 0.5)$ and the \emph{Kolmogorov-Smirnov test statistic}. 
While for the latter, R provides the function \code{ks.test()} one of the most popular ways to compute the AUC in R is given by the package \CRANpkg{ROCR} (\citealp{sing:2005, ROCR}). Nonetheless, for the purpose of credit scorecard modelling it is referred to the package \CRANpkg{pROC} at this point for the following three reasons:

\begin{enumerate}
	\item Different to standard binary classification problems credit scores are typically supposed to be increasing if the event (= default-) probability decreases. The function \code{roc()} of the package \pkg{pROC} has an argument \code{direction} which allows to specify this.
	\item In credit scoring applications it may be given that not all observations of a data set are of equal importance, e.g. it may be not as important to distinguish which of two customers with small default probabilities has the higher score if his or her application will be accepted anyway. The package's function \code{auc()} has an additional argument \code{partial.auc} in order to compute \emph{partial area under the curve} \citep{robin:2011}.
	\item Finally, its function \code{ci()} can be used to compute confidence intervals for the AUC using either bootstrap or the method of DeLong (\citealp{delong:1988, sun:2014}), e.g. in order to support the comparison of two models.	
\end{enumerate}
Example 16 demonstrates how \pkg{pROC} can be used for performance analysis.

\begin{example}
### example 16: Gini coefficient using {pROC} (based on example 13)
library(pROC)
curve <- roc(valid$creditability, valid_scored$score, 
             levels = c("good","bad"), direction = ">")
             # levels = c("controls", "cases"), 
             # direction = controls > cases 
plot(curve)
auc(curve)
# gini coefficient:
2 * (auc(curve) - 0.5)
# confidence limits for the auc:
ci(auc(curve), method = "bootstrap")
\end{example}

Among the packages enumerated above \pkg{creditR} offers a function \code{Kolmogorov.Smirnov()} and \pkg{riskr} has two functions \code{ks()} and \code{ks2()} for computation of the Kolmogorov-Smirnov test statistic. In addition, \pkg{riskr} provides a function \code{divergence()} to compute the divergence between two empirical distributions  as well as \code{gg\_dists()} and \code{gg\_cum()} to visualize the score densities for defaults and non-defaults and their empirical cumulative distribution functions. In order to compute the gini coefficient the package \pkg{riskr} provides functions \code{aucroc} (AUC), \code{gini} (Gini coefficient), \code{gg\_roc()} (visualization of the ROC curve), \code{gain()} (gains table for specified values on the x-axis) and \code{gg\_gain()} /\code{gg\_lift()} (for visualization of the gains-/lift-chart).  

The package \pkg{InformationValue} contains two functions \code{ks\_stat()} and \code{ks\_plot()} for Kolmogorov-Smirnov analysis and several functions: \code{AUROC()}, \code{plot\_ROC()}, \code{Concordance()} and \code{SomersD()} (Gini coefficient) to support analyses with regard to the Gini coefficient. 
Additionally, the \code{confusionMatrix()} and derivative performance measures \code{misClassError()}, \code{sensitivity()}, \code{specificity()}, \code{precision()}, \code{npv()},  \code{kappaCohen()} and \code{youdensIndex()} \citep[cf. e.g.][ch.5 for an overview]{zumel:2014} can be computed  for a given cut off by the corresponding functions. 
%Additional functions \code{confusionMatrix()}, \code{sensitivity()}, \code{specifity()}, \code{precision()}, \code{npv()}, \code{misClassError()}, \code{kappaCohen()} and \code{youdensIndex()} can be used to compute the corresponding measures for a given cut off \citep[cf. e.g.][ch.5]{zumel:2014}. 
Note that these measures are computed with respect to the non-default target level (supposed to be coded as '1' in the target variable) as well as a cut off optimization w.r.t. the missclassification error, Youden's Index or the minimum (/maximum) score such that no misclassified defaults (/non-defaults) do occur in the data (function \code{optimalCutoff()}).  
Similar measures (accuracy, precision, recall, sensitivity, specificity, F1) are computed by the function \code{fn\_conf\_mat()} of the \pkg{scorecardModelUtils} package. Numeric differences between the (0/1-coded) target and the model's predictions in terms of MSE, MAE and RMSE can be computed by its \code{fn\_error()} function.
The package \pkg{boottol} contains a function \code{boottol()} to compute bootstrap confidence intervals for Gini, AUC and KS where also subsets of the data above different cut off values are considered.
It may be desirable to analyze the (cumulative) frequencies of the binned scores. A table of such frequencies is returned by the function \code{gini\_table()} in the 
\pkg{scorecardModelUtils} package. Example 17 shows selected columns for a binned score using the function \code{gains\_table()} from the \pkg{scorecard} package. 

\begin{example}
### example 17: score bin frequencies (...for valid_scored from example 13) 
library(scorecard)
gt <- gains_table(valid_scored$score, valid$creditability, bin_num = 8)
gt[,c(2,4,5,6,7,8,10,11,12)]

          bin cum_count good cum_good bad cum_bad    badprob approval_rate cum_badprob
1:  [628,Inf)        37   37       37   0       0 0.00000000     0.1262799  0.00000000
2:  [575,628)        76   36       73   3       3 0.07692308     0.2593857  0.03947368
3:  [529,575)       112   34      107   2       5 0.05555556     0.3822526  0.04464286
4:  [492,529)       148   30      137   6      11 0.16666667     0.5051195  0.07432432
5:  [448,492)       185   26      163  11      22 0.29729730     0.6313993  0.11891892
6:  [399,448)       222   21      184  16      38 0.43243243     0.7576792  0.17117117
7:  [353,399)       257   14      198  21      59 0.60000000     0.8771331  0.22957198
8: [-Inf,353)       293    8      206  28      87 0.77777778     1.0000000  0.29692833
\end{example}

Note that although the Gini coefficient is generally bounded by -1 and 1 the value it can take for a specific model strongly depends on the discriminability of the data. For this reason it is suitable in order to compare performance on different models on the same data rather than comparing performance across different data sets. 
In consequence, for the purpose of an out-of-time monitoring of a scorecard it is thus rather advisable to compare an existing scorecard's performance against a recalibrated version of it than to compare it with its performance on the original (development) data. 
Drawbacks of the Gini coefficient as a performance measure for binary classification are discussed in \citet{hand:2009} and the \emph{H-measure} is proposed as an alternative which is implemented in the package \CRANpkg{hmeasure} \citep{hmeasure}. The \emph{expected maximum profit} measure \citep{verbraken:2014} as implemented in the package \CRANpkg{EMP} \citep{EMP} further takes into account for the profitability of a model.

\noindent {\bf Performance summary:} 

Many of the functionalities as provided by the packages for scorecard modelling in the previous subsection already exist in other packages and are thus not indispensable. 
In addition to these, however some of the packages are providing performance summary reports of several performance measures. These functions are listed in the following table:

\begin{table}[ht]
\centering
\footnotesize
\begin{tabular}{l|c|c|c|c}
  \toprule
  Package & \pkg{riskr} & \pkg{scorecard} & \pkg{scorecardModelUtils} & \pkg{smbinning} \\ 
  Function & \code{perf()} & \code{perf\_eva()} & \code{gini\_table()} & \code{smbinning.metrics()} \\ 
  \midrule
  KS & x & x & x & x \\ 
  AUC & x & x &  & x \\ 
  Gini & x & x & x &  \\ 
  \midrule
  Divergence & x &  &  &  \\ 
  Bin table &  &  & x &  \\ 
  Confusion matrix &  & x &  & x \\ 
  Accuracy &  &  &  & x \\ 
  Good rate &  &  &  & x \\ 
  Bad rate &  &  &  & x \\ 
  TPR &  &  &  & x \\ 
  FNR &  &  &  & x \\ 
  TNR &  &  &  & x \\ 
  PPV &  &  &  & x \\ 
  False discovery rate &  &  &  & x \\ 
  False omission rate &  &  &  & x \\ 
  NPV &  &  &  & x \\ 
  \hline
  ROC curve & x & x & x & x \\ 
  Score densities $|$ y & x &  &  &  \\ 
  ECDF & x & x &  & x \\ 
  Gain chart & x &  &  &  \\ 
   \bottomrule
\end{tabular}
\label{Tab:performance_summary}
\caption{Overview of scorecard performance summary functions.} 
\end{table}

In example 18 computation of a scorecard performance summary is demonstrated using the package \pkg{smbinning} (which returns the largest number of performance measures of the four functions from table 5) as well the function \code{riskr::gg\_perf()} that can be used to produce several graphs on the scorecard's performance (cf. Fig.~4). Note that although ROC curves are one of the most popular tools for performance visualization of binary classifiers they are hardly suited to visualize the performance difference of several competitive models. One reason for this is that large areas of the TPR-FPR plane (e.g. everything below the main diagonal) are typically of no interest given a specific data situation. For this reason in practice ROC curves are not very useful for model selection. 

\begin{example}
### example 18: scorecard performance summary (based on example 13)
library(smbinning)
perf_dat <- data.frame("creditability" = as.integer(valid$creditability == "good"), 
                       "score" = valid_scored$score)
smbinning.metrics(perf_dat, "score", "creditability", cutoff = 450) 

# roc curve, ecdf, score distribution and gain chart
library(riskr)
gg_perf(as.integer(valid$creditability == "good"), valid_scored$score)
\end{example}

\begin{figure}[htbp]
  \centering
  \includegraphics[width = 11cm]{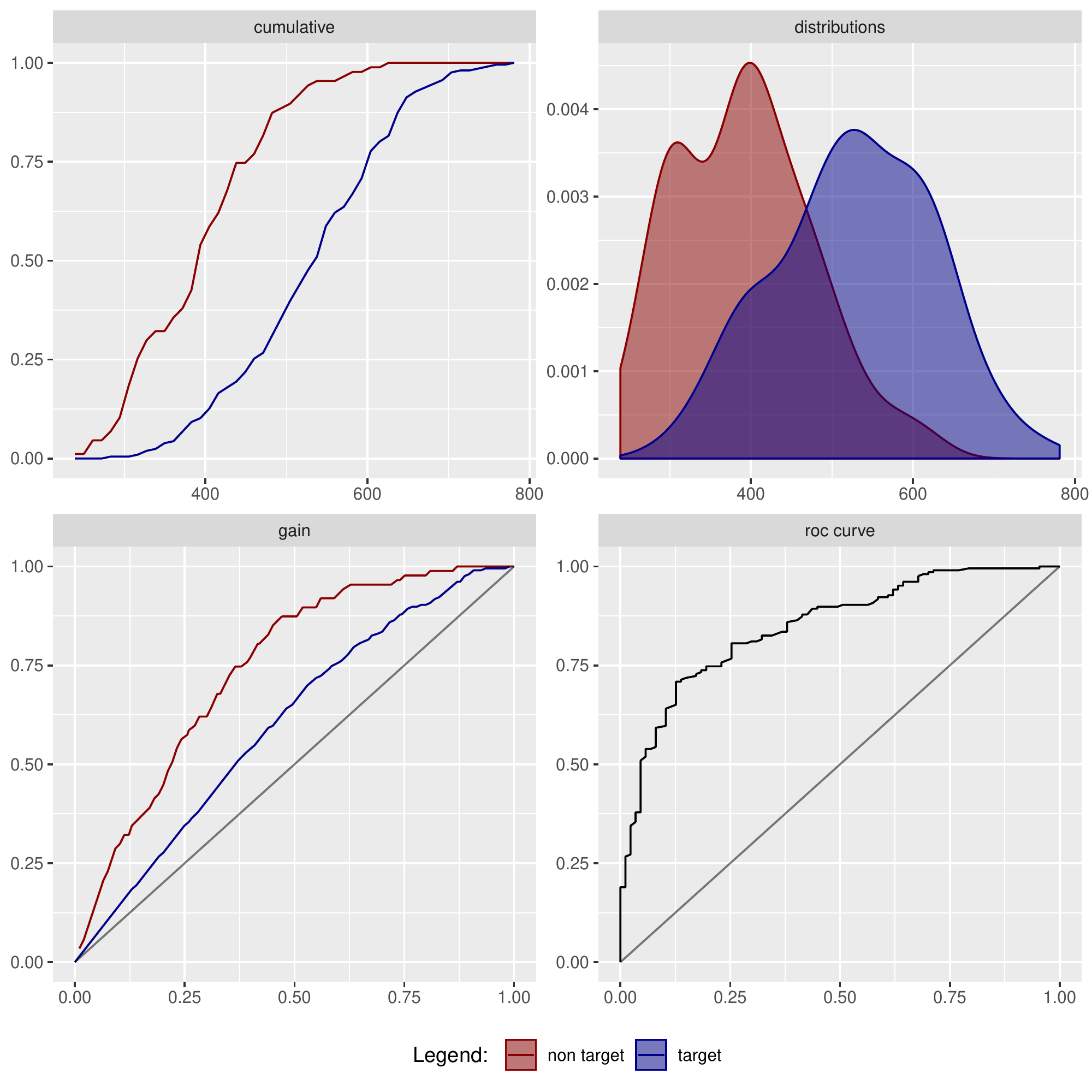}
  \caption{Scorecard performance graphs: ECDF (top left), score densities (top right), gains (bottom left) and ROC (bottom right).}
    \label{figure:bins}
\end{figure}

\noindent {\bf Rating calibration and concentration:} 

From a practical point of view it is often desirable to aggregate scorecard points into classes (\emph{rating grades}) of similar risk which is once again a binning task (cf. Sec.~\nameref{sec:binning}). The package \pkg{creditR} contains a function \code{master.scale()} that takes a data frame with scores and corresponding default probabilities as input and uses the function \code{woeBinning::woe.binning()} to group scores of similar WoE (cf. example 19). The function \code{odds\_table()} of the \pkg{riskr} package allows to set a \code{breaks} argument with arbitrary bins. 

Rating classes should be appropriately calibrated in the sense that the predicted and observed default probabilities do match for all rating grades. In order to check this the package \pkg{creditR} contains three functions  (\code{chisquare.test()}, \code{Binomial.test()} and \code{Adjusted.Binomial.test()}) that do provide a table with indicators for each rating grade (cf. example 19). Another function \code{Anchor.point()} compares the observed average predicted default probability on the data with pre-specified boundaries around some desired \emph{central tendency} default probability. Bootstrap confidence intervals for default probabilities of rating grades can be computed using the function \code{vas.test()} of the package \pkg{boottol}.
A Hosmer-Lemeshow goodness-of-fit test \citep{hosmer:2000} is e.g. implemented by the function \code{hoslem.test()} in the \CRANpkg{ResourceSelection} package \citep{ResourceSelection}. 

\pagebreak

\begin{example}
### example 19: rating calibration analysis (based on example 13)
library(creditR)
# calculate PDs from scores
odds <- 1/19 * 2^(-(valid_scored$score - 600)/50)
pd   <- odds / (1 + odds)
pd.dat <- data.frame(pd = pd, creditability = as.integer(valid$creditability == "bad"))
# aggregate scores to rating grades 
mscale <- creditR::master.scale(pd.dat, "creditability", "pd") 
# transform $Bad.Rate into numeric
mscale$Bad.Rate <- as.numeric(gsub("%","",mscale$Bad.Rate))/100
# test calibration of the rating grades
chisquare.test(mscale, "PD", "Bad.Count", "Total.Observations")
bintest <- Binomial.test(mscale, "Total.Observations", "PD", "Bad.Rate")

bintest[,c(1,2,3,4,5,8,9,14)]
\end{example}

\begin{example}
   Final.PD.Range Total.Distr Good.Count Bad.Count Bad.Rate      PD          Test_Result
1 <= 0.0692759267       25.9%         73         3    0.039 0.03835 Target Value Correct
2 <= 0.1477666759       17.4%         47         4    0.078 0.11149 Target Value Correct
3 <= 0.1904265313        7.2%         17         4    0.190 0.17482 Target Value Correct
4 <= 0.275937974         9.6%         19         9    0.321 0.23297 Target Value Correct
5 <= 0.3709582356        7.5%         14         8    0.364 0.32236 Target Value Correct
6 <= 0.4365863463        5.1%          8         7    0.467 0.41036 Target Value Correct
7 <= 0.4605851205        3.1%          6         3    0.333 0.45143 Target Value Correct
8 <= 0.5695614029        9.2%         10        17    0.630 0.51257 Target Value Correct
9          <= Inf       15.0%         12        32    0.727 0.72768 Target Value Correct
\end{example}

According to regulation ratings must avoid \emph{risk concentration} (i.e. a majority of the observations being assigned to only few grades). The Herfindahl-Hirschman index ($HHI = \sum_j \hat{f}(j)^2$, with the empirical distribution $\hat{f}$ of the rating grades $j$) can be considered to verify this as e.g.  implemented by \pkg{creditR}'s \code{Herfindahl.Hirschman.Index()} or \code{Adjusted.Herfindahl.Hirschman.Index()}. Small values of HHI mean that there is low risk concentration.  

\noindent {\bf Cross-validation:}

Some of the mentioned packages also provide functions for cross-validation. As both binning and variable selection are interactive they are not suited for cross-validation (cf. Sec. \nameref{sec:binning} and \nameref{sec:preselection}). For this reason it should rather be used on the training data and restricted to analyzing overfitting of the logistic regression model. 
There are already several packages available that do provide general functionalities for execution of cross-validation analyses (e.g. \pkg{mlr3} or \pkg{caret}). The function \code{k.fold.cross.validation.glm()} of the \pkg{creditR} package computes cross-validated Gini coefficients while the function \code{perf\_cv()} of the \pkg{scorecard} package offers an argument to specify different performance measures such as \code{"auc"}, \code{"gini"} and \code{"ks"}. 
Both functions do allow to set seeds in order to guarantee reproducibility of the results. 
The function \code{fn\_cross\_index()} somewhat more generally returns a list of training observation indices that can be used to implement an own cross-validation and compare models using identical folds.

\section{Reject inference}\label{sec:ri}
Typically, the final stage of a scorecard development consists in \emph{reject inference}: The scorecard model is based on historical data but already in the past, credit applications of customers that were assumed to be of high risk were rejected and thus for these data only the predictor variables are available from the application but not the target variable. The use of these observations with unknown performance is commonly referred to as reject inference. 

The benefits of using reject inference in practice still remains questionable. It has been investigated by several authors (cf. e.g. \citealp{crook:2004, banasik:2007, verstraeten:2005,  buecker:2013, ehrhardt:2019ri}) and is nicely discussed in \citet{hand:1993}. 
The appropriateness of different suggested algorithms for reject inference depends on the way how the probability of being rejected can be modelled, i.e. whether it is solely a function of the scorecard variables (MAR) or not (MNAR) \citep[for further details cf. also][]{little:2002}. 
A major issue is that especially for the most relevant MNAR situation the inference entirely relies on expert judgements and for this reason the appropriateness of the model can't be tested anymore. In consequence, reject inference should be used with care.

In R, the only package which offers functions for reject inference is the package \pkg{scoringTools} which is only available on Github but not on CRAN. It provides five functions for reject inference:
\code{augmentation()}, \code{fuzzy\_augmentation()}, \code{parcelling()}, \code{reclassification()} and \code{twins()} which do correspond to common reject inference strategies of the same name \citep[cf. e.g.][]{finlay:2012}. 
In the following, two of the most popular strategies, namely \emph{augmentation} and \emph{parcelling} are briefly explained as they are implemented within the package, completed by an example of their usage. 

\noindent {\bf Augmentation:}

%Augmentation is suitable for the MAR situation \citep{ehrhardt:2019ri}: 
An initial logistic regression model is trained on the observed data of approved credits (using all variables, i.e. variable selection has to be done in a preceeding step). Afterwards, weights are assigned to all observations of this sample of accepted credits, according to their probability of being approved: 
For this purpose, all observations (accepted and rejecetd) are scored by the initial model. Then, score-bands are defined and within each band\footnote{For the function \code{augmentation()} this is obtained by rounding the posterior probabilities to the first digit.} the probability of having been approved is computed by the proportion of observations with known performance in the combined sample from both accepted and rejected credits.
Finally, the logistic regression model is fitted again on the sample of the accepted loans only with observed performance but re-weighted observations\footnote{Here, the augmented weights within each score-band are computed by $1+\frac{n_{rejected}}{n_{accepted}}$.}.

\noindent {\bf Parcelling:}

Based on an initial logistic regression model which is trained on the observed data of approved credits only score-bands are defined and the observed default rate $\widehat{PD}_j$ of each score-band $j$ is derived. The observations of the rejected subsample are then scored by the initial model and assigned to each score-band. Labels are randomly assigned to the rejected observations such that they will have a default probability of $\widehat{PD}_j \times \alpha_j$\footnote{Within the function  \code{parcelling()} this is done by sampling the labels from a binomial distribution.} in each band 
where $\alpha_j$ are user-defined upweights of the score-bands' default rates which have to be specified by expert experience. 
Typically the $\alpha_j$ are set to be increasing for score-bands with larger default probabilities. 
Note that accepting these credit applications in the past might have happend for reasons beyond those that were reflected by the score variables but which led to a reduced risk for these observations in the observed sample compared to observations with a similar score in the total population. For this reason, parcelling is suitable for the MNAR situation. 

Example 20 illustrates parcelling using the \pkg{scoringTools} package. Note that all other functions of this package are of similar syntax and output. 
For parcelling in particular, the \code{probs} argument specifies quantiles w.r.t. the predicted default probabilities (i.e. from low risk to high risk). Although in the example the upweight vector \code{alpha} is constantly set to 1 for all bands, in practice it will rather be chosen to be increasing, at least for quantiles of high PDs.   

\begin{example}
### example 20: reject inference using parcelling (based on example 4) 
library(scoringTools)
# use validation data as 'rejects' for this example
# ...remove target variable and constant variable foreign.worker_bin
reject_woes <-  valid_woes[,-(1:2)]
# apply parcelling
set.seed(42) # reproducibility
ri_parc <- parcelling(xf = train_woes[,-(1:2)], xnf = reject_woes, 
                      yf = ifelse(train_woes[,1] == "bad", 1, 0), 
                      probs = c(0, 0.25, 0.5, 0.7, 0.8, 0.9, 1), alpha = rep(1, 6))
# final model after reject inference
class(ri_parc@infered_model)   
# observations weights
ri_parc@infered_model$weights
# combined sample after parcelling (note automatically renamed variables)
str(ri_parc@infered_model$data)

# recompute WoEs on combined sample using weight (cf. also example 4)
combined_bins                <- rbind(train_bins, valid_bins)
combined_bins$creditability  <- as.factor(ifelse(ri_parc@infered_model$data$labels == 1, 
                                                 "bad", "good")) 
library(klaR)
woe_model_after_ri <- woe(creditability ~ ., data = combined_bins, 
                          weights = ri_parc@infered_model$weights)
combined_woes      <- data.frame(creditability = combined_bins$creditability, 
                                 woe_model_after_ri$xnew)
\end{example}

The initial model and the final model are stored in the result object's slots \code{financed\_model} and \code{infered\_model}. Both are of class \code{glm}. Note that both models are automatically calculated without any further options of parameterization such as e.g. variable selection or a re-computation of the WoEs based on the combined sample of accepted applications and rejected applications with inferred target. 
For this purpose the \code{woe()} function of the \pkg{klaR} package can be used which supports the 
specification of observations weights as the only one among all presented packages.
Finally, the combined sample can be used to rebuild the scorecard model as described in Sections \nameref{sec:preselection}, \nameref{sec:scorecard} and \nameref{sec:performance}.

\section{Summary}\label{sec:summary}
For a long time in the {R} universe no packages were available that were explicitly dedicated to the credit risk scorecard development process while during the last five years a simultaneous growth of several packages on this task has been observable. Some of these packages are available on CRAN while some are only available on Github.
 
The paper aims to give a comparative overview on the different functionalities of all currently available packages guided by the sequence of steps along a typical scorecard development process. In conclusion, inbetween any required functionality is available making it easy to develop scorecards using \R. According to the author's personal opinion, currently the most comprehensive implementations are given by the packages \pkg{scorecard}, \pkg{scorcardModelUtils} and \pkg{smbinning} and the package \pkg{woeBinning} provides an implementation of binning and WoE computation that reflects a broad range of practical issues (cf. Sec. \nameref{sec:binning}).
Grace to its large developing community and the huge amount of freely available packages developers have access to many additional packages which are not explicitly designed for the purpose under investigation but still provide valuable tools and functions to facilitate and improve the analyst's life making {R} a serious alternative to commercial software on this topic. 

Investigation of the functionalities provided by the different packages turns out that the packages seem to have been developed quite independently of each other: 
Some steps of the developments are addressed in many packages, especially the important one of binning variables but links between the packages are mostly missing\footnote{As an exception, the package \pkg{creditR} has been developed as an extension of the package \pkg{woeBinning} and the \pkg{woeR} package uses data from \pkg{smbinning} in the examples.} and many packages are not flexibly designed in the sense that their functions do require input arguments and variable naming conventions restricted to results from functions of the same package which makes it somewhat difficult to benefit from advantages of different packages at the same time. The paper's supplementary code provides some remedies for this difficulty\footnote{Cf. corresponding footnotes in the paper. All codes are available under \url{https://github.com/g-rho/CSwR}.}. Some of the packages are even missing predict functionalities in order to apply the results of the modelling to new data. 
In order to summarize the results as they have been worked out in the previous sections the following table lists the presented packages with an explicit scope of scorecard modelling 
and the stages of the development process that are adressed by them:

\begin{table}[ht]
\centering
\footnotesize
\begin{tabular}{l|ccccc}
  \hline
Package & Binning \& WoEs & Preselection & Scorecard & Performance & Reject Inference \\ 
  \toprule
boottol &  &  &  & x &  \\ 
  creditR & x & x & x & x &  \\ 
  glmdisc & x &  & x &  &  \\ 
  glmtree &  &  & x &  &  \\ 
  Information &  & x &  &  &  \\ 
  InformationValue &  & x &  &  &  \\ 
  riskr & x & x &  & x &  \\ 
  Rprofet & x &  & x &  &  \\ 
  scorecard & x & x & x & x &  \\ 
  scoringTools & x &  &  &  & x \\ 
  scorecardModelUtils & x & x & x & x &  \\ 
  smbinning & x & x & x & x &  \\ 
  woe & x & x &  &  &  \\ 
  woeBinning & x &  &  &  &  \\ 
  woeR & x &  &  &  &  \\ 
   \bottomrule
\end{tabular}
\label{Tab:summary_pkgs_stages}
\caption{Overview of R packages with the explicit scope of scorecard modelling and addressed stages of the development process.} 
\end{table}

Finally, and with regard to the title of the paper a the last figure aims to visualize the 'landscape' of R packages dedicated to scorecard development using \emph{logistic principal component analysis} \citep{landgraf:2015} as implemented in the \CRANpkg{logisticPCA} package \citep{logisticPCA} on the binary data as given by table 6. 

\pagebreak

\begin{figure}[htbp]
  \centering
  \includegraphics[width = 12cm]{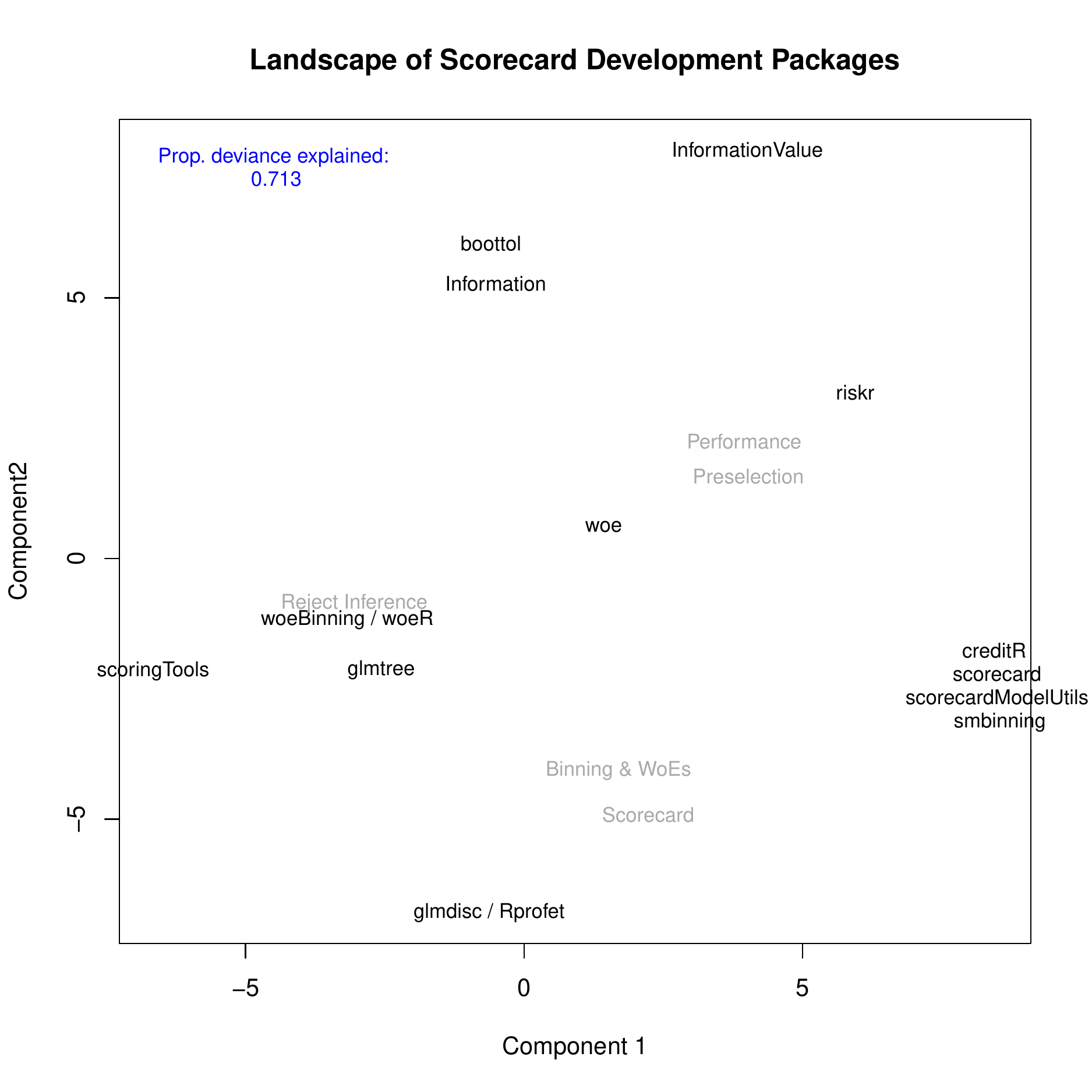}
  \caption{Landscape of R packages for scorecard modelling using logistic PCA.}
    \label{figure:pkgstats}
\end{figure}

The future will show up to what degree the traditional process of credit risk scorecard development will stay as it is or whether %and if it will then 
or up to what extent the use of logistic regression will be replaced by more recent machine learning algorithms such as e.g. offered by the recent powerful \pkg{mlr3} framework in combination with explainable ML methodology to fulfill regulatory requirements.

\section{Acknowledgment}

Thanks to Rabea Aschenbruck and Alexander Frahm at Stralsund University of Applied Sciences for supporting me in writing this paper.

\bibliography{szepannek}

\end{article}

\end{document}